\documentstyle[12pt]{article}

\begin{document}
\setlength{\textheight}{1.2\textheight}


\def\beqn{\begin{equation}}
\def\eeqn{\end{equation}}
\def\beqa{\begin{eqnarray}}
\def\eeqa{\end{eqnarray}}


\def\q{\dot q}
\def\dq{\delta q}
\def\dvphi{\dot \varphi}
\def\dpsi{\dot \psi}

\begin{titlepage}
\nopagebreak
\begin{center}
{\large\bf QUANTUM FLUCTUATIONS AND INERTIA}\\

 \vfill
        {\bf Marc-Thierry Jaekel\dag ~and Serge Reynaud\ddag }  \\
\end{center}
\dag Laboratoire de Physique Th\'eorique
 de l'Ecole Normale  Sup\'erieure\footnote{Unit\'e
propre du Centre National de la Recherche Scientifique, \\
associ\'ee \`a l'Ecole Normale Sup\'erieure et \`a l'Universit\'e
de Paris Sud.}(CNRS),
24 rue Lhomond, F75231 Paris Cedex 05, France \\
\ddag Laboratoire Kastler-Brossel\footnote{Unit\'e
de l'Ecole Normale Sup\'erieure et de l'Universit\'e Pierre et Marie Curie,\\
 associ\'ee au Centre National de la Recherche
Scientifique.}(UPMC-ENS-CNRS), case 74,\\
 4 place Jussieu,  F75252 Paris Cedex 05,
France \\
\vfill

\vfill

\begin{abstract}
Vacuum field fluctuations exert a radiation pressure which induces
mechanical effects on scatterers. The question naturally arises
whether the energy of vacuum fluctuations gives rise
to inertia and gravitation in agreement with the general principles
of mechanics.  As a new approach to this question, we discuss the mechanical
effects of quantum field fluctuations on two
mirrors building a Fabry-Perot cavity.  We first put into evidence that
the energy related to Casimir forces is an energy stored on field
fluctuations as a result of scattering time delays. We then discuss
the forces felt by the mirrors when they move
within vacuum field fluctuations, and show that energy stored on vacuum
fluctuations contributes to inertia in conformity with the law
of inertia of energy. As a further consequence, inertial masses exhibit quantum
fluctuations with characteristic spectra in vacuum.

\end{abstract}
\vfill

\begin{flushleft}

\vfill
        {\normalsize LPTENS 95/4\\
May 1995\\}

\vfill

Proceedings of NATO-ASI Conference "Electron Theory and Quantum Electrodynamics
- 100
years later", September 5-16, 1994, Edirne, Turkey.

\end{flushleft}

\end{titlepage}

\begin{flushleft}
{\bf 1 Introduction}
\end{flushleft}

Fundamental problems are raised by the mechanical effects associated with
radiation pressure fluctuations in vacuum. The
instability of motions when radiation reaction is taken into account,
and the existence of "runaway solutions" \cite{Rohrlich},
 can be avoided for mirrors by recalling
that they are actually transparent to high frequencies
of the field \cite{JR4}. However, partially transmitting mirrors, and
cavities, introduce
scattering time delays which result in a temporary storage
of part of the scattered vacuum fluctuations \cite{JR1}.
In particular, the energy related to Casimir forces \cite{Casimir}
identifies with the energy of field fluctuations stored in the cavity
\cite{JR1}.
This revives the questions of the contribution of vacuum fluctuations
to inertia and gravitation \cite{Nernst},
and of its consistency with the general principles
of equivalence and of inertia of energy.

Vacuum fluctuations of quantum fields have been known for long to correspond to
an infinite energy density \cite{Enz}, or at least to a problematically high
energy
density, if only frequencies below Planck frequency
are considered \cite{Wesson}.
A common way to escape the problems raised by consequent
gravitational effects, exploits the fact that only differences of energy
are involved in all other interactions. Vacuum energy is set to zero
by definition, a prescription which is embodied in normal ordering of
quantum fields. In such a scheme, variations of vacuum energy, like the
energy associated with Casimir forces \cite{Casimir}, hardly give rise to
inertia
and gravitation. Furthermore, normal ordering cannot be implemented as a
covariant prescription and leads to ambiguities in defining the
gravitational effects of quantum fields \cite{BD}.
Then, the question naturally arises of the compatibility
of the mechanical effects induced by quantum field fluctuations with
the general principles which govern the laws of mechanics.

As a new approach to this question, we discuss the mechanical
effects of quantum field fluctuations on two
mirrors building a Fabry-Perot cavity.  We first put into evidence that
the energy related with Casimir forces is an energy stored on field
fluctuations as a result of scattering time delays. We then discuss
the forces felt by the mirrors when they move
within vacuum field fluctuations, and in particular the contribution
of Casimir energy to inertia.

\bigskip
\begin{flushleft}
{\bf 2 Casimir energy}
\end{flushleft}

As a result of the radiation pressure of field quantum fluctuations
in which they are immersed, two mirrors at rest in vacuum feel a mean
Casimir force which depends on their distance $q$.
For partially transmitting mirrors,
characterised by their frequency dependent reflection coefficients
($r_1$ and $r_2$), the Casimir force takes a simple form
 (written here for a cavity
in two-dimensional space-time immersed in the vacuum of a scalar field;
similar expressions hold in four-dimensional space-time, and for
electromagnetic and also thermal fields) \cite{JR1}:
$$F_c = \int_0^\infty {d\omega \over 2\pi} {\hbar \omega \over c}
\lbrace 1 - g[\omega]\rbrace$$
$$g[\omega] = {1 - |r[\omega]|^2 \over |1 - r[\omega]e^{2i\omega q/c}|^2}
\qquad \qquad r[\omega] = r_1[\omega] r_2[\omega]$$
The first part of this expression corresponds to the energy-momentum
of incoming vacuum field fluctuations ($\hbar$ is Planck constant, and $c$
the light velocity).
The second part describes the effect of the cavity on the modes:
$g[\omega]$ describes an enhancement of energy density for modes inside the
resonance peaks of the cavity, and an attenuation for modes outside.

This mean force can be seen as the variation of a potential energy,
more precisely, as the length dependent part of the energy of the cavity
immersed in field fluctuations:
$$dE_c = F_c dq$$
One easily derives the well-known phase-shift representation of
Casimir energy \cite{Casimir},
whose expression in the present case takes the simple
following form:
$$E_c = \int_0^\infty {d\omega \over 2\pi} \hbar
\lbrace-\delta[\omega]\rbrace$$
$$2 \delta[\omega] = i Log {1 - r[\omega]e^{2i\omega q/c} \over
1 - r[\omega]^* e^{-2i\omega q/c}}$$
$$det S = det S_1 det S_2 e^{2i\delta}$$
$\delta[\omega]$ is the frequency dependent phase-shift introduced by the
cavity on the propagation of field modes, as given by the
scattering matrix ($S$) of the cavity
 (more precisely, its definition divides by
the individual scattering matrices of the mirrors, whose contributions
to the total energy are length independent).

The frequency dependent phase-shift corresponds to time delays in the
propagation of fields through the cavity:
$$\tau[\omega] = \partial_\omega \delta[\omega]$$
This time delay \cite{Wigner} describes the time lag undergone by a wave
packet around frequency $\omega$ and is the sum of several contributions:
\beqa
\label{td}
\tau[\omega] &=& - \lbrace 1 - g[\omega] \rbrace \lbrace {q\over c} +
{1\over 2}\partial_\omega \varphi\rbrace \nonumber\\
&+& g[\omega] sin( 2\omega {q\over c} + \varphi){ \partial_\omega
\rho \over 1-\rho^2} \nonumber\\
r[\omega] &=& \rho[\omega] e^{i\varphi[\omega]}
\eeqa
The main contribution identifies with the length of the cavity (divided by
$c$),
modified by the function $g$ describing energy
densities within the cavity. Other contributions are corrections due to the
frequency dependence of the mirrors' reflection coefficients, i.e. delays
introduced during reflection on the mirrors themselves.

Casimir energy can be rewritten in terms of these scattering time delays,
integrating by parts and noting that boundary terms vanish in particular
because of high frequency transparency:
$$E_c = \int_0^\infty {d\omega \over 2\pi} \hbar \omega \tau[\omega]$$
The result takes a simple form, as an integral over all modes of the product
of the spectral energy density of quantum field fluctuations by the
corresponding time delay. In particular, the length dependent part of
Casimir energy is negative, corresponding to a binding energy,
so that negative time delays contribute in majority \cite{BCG}. As time delays
are
indeed relative to free propagation, i.e in abscence of cavity, the retardation
effect of the cavity on resonant modes is thus dominated  by the opposite
effect on modes outside resonance peaks.

It can be shown that the same expressions remain valid for Casimir force and
energy of a cavity immersed in thermal fields, provided the spectral energy
density for thermal quantum fluctuations is substituted ($T$ is the
temperature)
\cite{JR1}:
$$F_c = \int_0^\infty {d\omega \over 2\pi }2 {\hbar \omega \over c}
\lbrace {1\over 2} + {1\over e^{\hbar \omega /T} - 1} \rbrace
\lbrace 1 - g[\omega] \rbrace $$

 $$E_c = \int_0^\infty {d\omega \over 2\pi}2 \hbar \omega \lbrace
{1\over 2} + {1\over e^{\hbar \omega /T} - 1} \rbrace \tau[\omega]$$
To the contribution of zero-point fluctuations, one must add the contribution
due to the mean number of photons as given by Planck's formula.
In all cases, Casimir energy appears as part of the energy of quantum
field fluctuations which is stored inside the cavity, as a consequence
of scattering time delays.

\bigskip
\begin{flushleft}
{\bf 3 Motional Casimir forces}
\end{flushleft}

The Casimir forces felt by two mirrors at rest result from the radiation
pressure exerted by the fluctuating quantum fields in which they are
immersed. Hence, these forces also fluctuate and their fluctuations can be
characterised by their correlations ($i,j = 1,2$ label the two mirrors):
$$<F_i(t) F_j(0)> - <F_i><F_j> = C_{F_iF_j}(t)$$
For a stationary state of the field, correlations are equivalently
characterised by spectral functions \cite{JR3}:
\beqn
\label{ff}
C_{F_iF_j}(t) = \int_{-\infty}^\infty {d\omega \over 2\pi} e^{-i\omega t}
C_{F_iF_j}[\omega]
\eeqn
The fluctuating forces induce random motions of the mirrors around their mean
positions which can be described as quantum Brownian motions.
As a consequence of general principles governing motion in
a fluctuating environment \cite{E1}, when set into motion
mirrors feel additional forces which depend on their motions. For small
displacements ($\dq_i$), these forces are conveniently described
by motional susceptibilities:
\beqn
\label{mf}
<\delta F_i[\omega]> = \sum_j \chi_{F_iF_j}[\omega] \dq_j[\omega]
\eeqn
The motional forces can be obtained using
motion dependent scattering matrices \cite{JR2}.
The scattering matrix of a mirror in its rest frame leads to a scattering
matrix in the original frame which depends on the mirror's motion, and
can easily be obtained up to first order in the mirror's displacement.
Radiation pressures and forces exerted on the mirrors are thus obtained
up to the same order \cite{JR3}. (For perfect mirrors, forces have been
obtained exactly for arbitrary motions of the mirrors \cite{Moore}).

According to linear response theory \cite{Kubo},
fluctuation-dissipation relations identify the imaginary (or dissipative)
part of a susceptibility with the commutator of the corresponding
quantity with the generator of the perturbation.
In the case of mirrors' displacements, the generators are the forces
exerted on the mirrors:
$$\chi_{F_iF_j}[\omega] - \chi_{F_jF_i}[-\omega] =
{i \over \hbar}\lbrace C_{F_iF_j}[\omega] -  C_{F_jF_i}[-\omega]\rbrace$$
Thus, fluctuation-dissipation relations provide a check for the results
one obtains independently for force fluctuations (\ref{ff}) and
for motional susceptibilities (\ref{mf}).

Although rather complex in their total generality, explicit expressions
for motional forces induced by vacuum fluctuations on partially transmitting
mirrors satisfy some general interesting properties \cite{JR3}.
As expected, the
motional forces present mechanical resonances for frequencies which
coincide with optical modes of the cavity:
$$\omega = n \pi {c \over q}$$
Although motional Casimir forces are naturally small, much smaller than
static Casimir forces, resonance properties might be used to compensate
their smallness using cavities with very high quality factors, thus
possibly leading to experimental evidence.

Other interesting
properties of these forces appear at the quasistatic limit, i.e. at
the limit of very slow motions \cite{JR8}. For displacements which vary slowly
in time, one can use a quasistatic expansion (expansion around zero
frequency $\omega \sim 0$) of the expressions for motional
susceptibilities (\ref{mf}) (a dot stands for time derivative):
$$<\delta F_i[\omega]> = \sum_j \lbrace \chi_{F_iF_j}[0] \dq_j[\omega] +
{1\over 2}\chi_{F_iF_j}^{''}[0] \omega^2 \dq_j[\omega] + \ldots\rbrace$$
$$<\delta F_i(t)> = -\sum_j \lbrace \kappa_{ij} \dq_j(t) +
\mu_{ij} \delta{\ddot q}_j(t) + \ldots\rbrace$$
The first term, described by $\kappa_{ij}$ ($ - \chi_{F_iF_j}[0]$), just
reproduces the variations of the static Casimir force when
the length of the cavity is changed. The further terms correspond
to new forces which emerge when the mirrors are accelerated in vacuum
and which exhibit peculiar features.
These forces are proportional to the mirrors' accelerations and are
conveniently expressed under the form of a mass matrix
$\mu_{ij}$ ($ {1\over 2}\chi_{F_iF_j}^{''}[0]$).
 Diagonal terms are corrections to the mirrors' masses. They
show that each mirror's mass is modified by the presence of the other mirror,
with a correction which depends on the distance between the two mirrors.
But non diagonal terms are also present, corresponding to the emergence of an
inertial force for one mirror when the other mirror is accelerated.
These properties of the inertial forces induced by vacuum fluctuations
are reminiscent of Mach's principle of relativity of inertia.
They indeed satisfy the requirements that Einstein \cite{E3}
stated in his analysis
of Mach's conception of inertia and in the context of gravity.
They strongly suggest a relation between modifications of vacuum fields
and gravitational effects \cite{Sakharov}.

\bigskip
Inertial forces acting on the cavity as a whole are
related with global motions of the cavity,
i.e. identical motions of the two mirrors (in linear approximation for
displacements):
$$\delta{\ddot q}_1(t) = \delta{\ddot q}_2(t) = \delta{\ddot q}(t)$$
The total force acting on the cavity moving in vacuum fields then contains a
component which dominates for slow motions and which is proportional to
the cavity's acceleration:
$$<\delta F(t)> = <\delta F_1(t) + \delta F_2(t)> =
-\lbrace \mu \delta{\ddot q}(t) + \ldots\rbrace$$
$$\mu = \sum_{ij} \mu_{ij}$$
Explicit computation \cite{JR8} shows that
 the corresponding mass correction for the
cavity is proportional to
the length of the cavity and to the Casimir force between the two mirrors:
\beqn
\label{cmc}
\mu c^2 = - 2 F_c q
\eeqn
This correction appears to be proportional to the contribution of the
intracavity fields to the Casimir energy, i.e the energy stored on vacuum
fluctuations due to the propagation delay inside the cavity
(see (\ref{td})). For a cavity built with perfect
mirrors in particular, this corresponds to Casimir energy:
$$E_c = - F_c q$$
Although not quite obvious at first sight, the factor $2$ is in fact the
correct one in the present case. Indeed, it was already shown by Einstein
\cite{E2}, that for a stressed rigid body Lorentz invariance implies a
relation for the mass ($\mu$), i.e the ratio between momentum and velocity,
that not only involves the internal energy of the body ($E_c$) but also
the stress ($F_c$) exerted on the body:
$$\mu c^2 =  E_c - F_c q$$
When comparing the total momentum with the velocity of
the center of inertia of the whole system,
i.e. taking into account not only the masses of the two mirrors but also
the energy stored in the fields inside the cavity, this relation leads to the
usual equivalence between mass and energy. Thus,
the energy of vacuum field fluctuations stored inside the cavity contributes to
inertia in conformity with the law of inertia of energy.

However, for partially transmitting mirrors, the energy stored according to
time delays due to reflection upon the mirrors (see (\ref{td})) is missing in
the
mass correction (\ref{cmc}). The inertial forces obtained for a cavity
moving in vacuum satisfy the law of inertia of energy for the energy of vacuum
fluctuations stored inside the cavity, but not for the energy stored in the
mirrors themselves. This result must be compared with a previous computation
of the force exerted on a single, partially transmitting, mirror moving
in vacuum fields, which appeared to vanish for uniformly
accelerated motion \cite{JR2}. This discrepancy with the general
equivalence between mass and energy reflects a defect in the representation
of the interaction of the mirror with the field.
We shall now discuss, using an explicit model
of interaction between mirror and field, how this representation can be
improved.

\bigskip
\begin{flushleft}
{\bf 4 Model of a pointlike scatterer}
\end{flushleft}

We consider the case of a scalar field $\phi$ interacting with a pointlike
mirror, located at $q$, in two-dimensional space-time ($(x^\mu)_{\mu=0,1} =
(t,x)$), described
by the following manifestly relativistic Lagrangian (from now on, $c=1$)
\cite{JR9}:
\beqa
\label{L}
{\cal A}  &=& \int {1\over2} (\partial\phi)^2 d^2x -
\int m \sqrt{1 - \q^2} dt\nonumber\\
{\cal L}  &=& {1\over2} (\partial\phi)^2 - m \sqrt{1 - \q^2}
\delta(x - q)
\eeqa
\beqn
\label{mc}
m = m_b + \Omega \phi(q)^2
\eeqn
The two terms are the usual Lagrangians for a free scalar field and a
free particle, except that the mass of the particle is assumed to also
contain a contribution which depends on the field. Such contribution
generally describes a relativistically invariant interaction term for the
field and the sources which are located on the mirror. In order to facilitate
comparison with
the simplified representation in terms of a $2\times2$ scattering matrix,
the interaction is further assumed to be quadratic in the field.
$\Omega$ is the inverse of a proper time characterising field scattering.
Equations for the field involve the scatterer's position and result in highly
non linear coupling:
\beqn
\label{fem}
\partial^2\phi = - 2\sqrt{1 - \q^2} \Omega \phi \delta(x - q)
\eeqn
However, if one considers as a first approximation that the mirror
remains at rest at a fixed position $q$,
then (\ref{fem}) becomes a linear equation describing
propagation in presence of a pointlike source. The field on both
sides of the scatterer decomposes on two components which propagate freely
in opposite
directions and which can be identified with incoming and outcoming fields.
The scattering matrix which relates outcoming and incoming modes is
obtained from equation (\ref{fem}), and identifies with a simple
symmetric $2\times2$ matrix determined by the following frequency dependent
diagonal ($s[\omega]$) and
non diagonal ($r[\omega]$) elements:
\beqn
\label{sm}
s[\omega] = 1 + r[\omega] \qquad \qquad
r[\omega] = - {\Omega \over \Omega - i\omega}
\eeqn
This corresponds to the simple model of partially transmitting mirror,
with a reflection time delay having a Lorentzian frequency dependence:
\beqn
\label{ltd}
\tau[\omega] = {\Omega \over \Omega^2 + \omega^2}
\eeqn
Simple computation shows that the energy stored on field fluctuations
due to this reflection time delay indeed identifies with the mass
term describing the interaction with the field (\ref{mc}). The mean mass
is determined by the correlations of the local field, which can be expressed
in terms of incoming correlations and of the scattering matrix. For incoming
fields in vacuum:
\beqn
\label{sse}
< \Omega \phi(q)^2 > = \int_0^\infty {d\omega \over 2\pi}
 \hbar \omega \tau[\omega]
\eeqn
Actually, the expression thus obtained for the mean value of the scatterer's
mass in vacuum is infinite, as a result of a diverging contribution of
high frequency fluctuations. In fact, the approximation
of a scatterer staying at rest, on which expression (\ref{ltd}) for the
time delay relies, cannot remain valid for sufficiently high frequencies.
At field frequencies which become comparable with the scatterer's mass,
recoil of the scatterer cannot be neglected, so that the simplified
$2\times2$ scattering matrix and its associated reflection
time delay fail to be good approximations.
Although consistent with the approximation which neglects
the scatterer's recoil for all field frequencies,
the result of an infinite stored energy does not correspond to the
general case.

 For a finite mass scatterer, the scatterer's recoil must be
taken into account. This is described by the equations of motion
for the scatterer which are derived from Lagrangian (\ref{L}):
\beqa
\label{rec}
{dp^\mu \over dt} &=& F^\mu = 2\Omega\sqrt{1-\q^2} \phi \partial^\mu \phi (q)
\nonumber\\
p^\mu &=& ({m \over \sqrt{1-\q^2}}, {m \q \over \sqrt{1-\q^2}})
\eeqa
These correspond to Newton equation, with a force depending on the
local field. Recalling the equations of motion for the field (\ref{fem}), the
force identifies with the radiation pressure exerted by the scattered field.
An important feature of the equations characterising the scatterer's recoil
is that the mass involved in the relation between the force and the scatterer's
acceleration includes the mass correction (\ref{mc}), that is
the energy stored by the scatterer on incoming field fluctuations.
As exemplified by this simple model, a correct treatment of the interaction
between field and a partially transmitting mirror
leads to an energy stored on vacuum field fluctuations due to reflection
time delays which also satisfies the universal equivalence between mass
and energy.

As shown by equations (\ref{rec}), the energy and momentum of the scatterer
satisfy the usual relations:
$$p_0^2 - p_1^2 = m^2 \qquad \qquad p^1 = p^0 \q$$
When submitted to the fluctuating radiation pressure of the field,
the scatterer undergoes a relativistic stochastic process which
remains causal, i.e with a velocity never exceeding the light velocity.
When fields with frequencies much smaller than the scatterer's mass
($\hbar \omega \ll <m>$) are reflected, recoil can be neglected and the
scattering matrix is well approximated by the linear $2\times2$
matrix (\ref{sm}). However, for frequencies of the order of the scatterer's
mass, recoil must be taken into account and the frequency dependence of
scattering time delays
differs significantly from the dependence at low frequencies (\ref{ltd}).
A complete and accurate treatment should then
consistently provide a finite stored
energy for a finite mass scatterer.

Integration of the stored energy in the inertial mass in a consistent way leads
to interesting new consequences. It directly results from their expressions
in terms of quantum field fluctuations (for instance
(\ref{sse})), that stored energies not only possess a mean value but also
fluctuations. Hence, the inertial mass is a fluctuating quantity, with
a characteristic noise spectrum:
$$<m(t)m(0)> - <m>^2 = \int {d\omega \over 2\pi} e^{-i\omega t}
C_{mm}[\omega] $$
For the pointlike scatterer just described, the inertial mass correction is
quadratic in the local field, and mass fluctuations are derived from
incoming field fluctuations and the scattering matrix. For frequencies well
below the scatterer's mass, recoil can be neglected and the mass noise spectrum
in vacuum is readily obtained form (\ref{sm}):
$$C_{mm}[\omega] = 2 \hbar^2 \theta(\omega)
\int_0^\omega {d\omega' \over 2\pi} \omega' \tau[\omega']
(\omega-\omega') \tau[\omega-\omega'] \qquad \qquad
(\hbar \omega \ll <m>)$$
This spectrum shows the characteristic positive frequency domain of vacuum
fluctuations. It also corresponds to a convolution (a direct
product in time domain)
of two expressions equal to the mean mass correction, a consequence of the
gaussian property of local field fluctuations (at this level of
approximation).  Inertial mass thus exhibits properties of a quantum variable.

As expected, mass fluctuations become extremely small for ordinary
time scales, i.e. for low frequencies. For frequencies below the
reflection cut-off $\Omega$, the mass noise spectrum grows like $\omega^3$:
$$C_{mm}[\omega] \simeq {\hbar^2  \over
6 \pi \Omega^2 } \theta(\omega) \omega^3 $$
The inertial mass remains practically constant in usual
mechanical situations. For high frequencies however, mass fluctuations become
important and cannot be neglected at very short time scales. As an illustration
(of course, recoil should be accounted for at such frequencies),
the same expression
exhibits mass fluctuations which become comparable with the mean mass
(for $m_b = 0$):
$$C_{mm}(t=0) = <m^2> - <m>^2 = 2 <m>^2$$

\bigskip
\begin{flushleft}
{\bf 5 Conclusion}
\end{flushleft}

Scattering time delays lead to a temporary storage of quantum field
fluctuations
by scatterers.
Vacuum quantum field fluctuations induce stored energies
and inertial masses which satisfy the universal equivalence between mass
and energy, including for their fluctuations. Vacuum fluctuations
result in mechanical effects which conform with general principles
of mechanics.
It can be expected that energies stored on quantum field fluctuations
should also lead to gravitation,
in conformity with the principle of equivalence.
Moreover, mass fluctuations due to vacuum field fluctuations
could play a significant role in a complete and consistent formulation
of gravitational effects.

\end{document}